\documentclass{moriond}

\usepackage{color,marvosym, pdfsync, colortbl}\usepackage{ulem}
\usepackage{graphicx, amssymb, epstopdf,url, hyperref,wrapfig,amsfonts,amsmath}

\usepackage[dvipsnames]{xcolor}

\def\arXiv#1{\href{http://arxiv.org/abs/#1}{arXiv:#1}} 

\newcommand{\TeV}{\,{\rm TeV}}
\newcommand{\MeV}{\,{\rm MeV}}
\newcommand{\GeV}{\,{\rm GeV}}
\newcommand{\eV}{\,{\rm eV}}
\def\circa#1{\,\raise.3ex\hbox{$#1$\kern-.75em\lower1ex\hbox{$\sim$}}\,}

\newcommand{\SU}{\,{\rm SU}}
\def\mathscr#1{{\cal #1}}
\newcommand{\pagina}{\subsection}
\newcommand{\beq}{\begin{equation}}
\newcommand{\eeq}{\end{equation}}
\usepackage{url}
\def\be{\begin{equation}}
\def\ee{\end{equation}}
\def\bea{\begin{eqnarray}}
\def\eea{\end{eqnarray}}

\newcommand{\Photo}{\includegraphics[height=35mm]{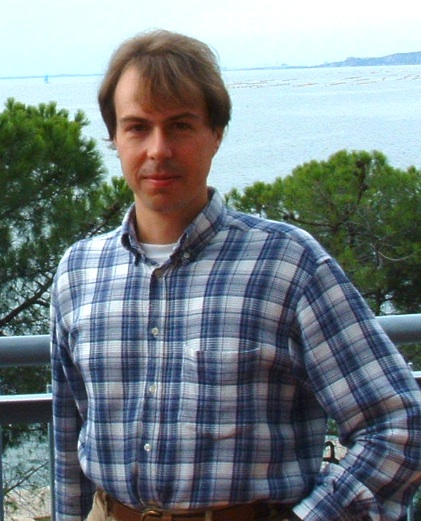}}

\begin{document}
\vspace*{4cm}
\title{THEORY SUMMARY OF MORIOND ELECTRO-WEAK 2015}

\author{ ALESSANDRO STRUMIA }

\address{Dipartimento di Fisica, Univ.\ di Pisa and INFN and NICPB}

\maketitle\abstracts{I summarise the theoretical talks at Moriond 2015,
with emphasis on naturalness.}


%
%
%

%


The theoretical talks at the electro-weak session of the
50th edition of the Moriond conference
touched many hot topics: neutrinos (section~\ref{nu}), dark matter (section~\ref{DMsec}), the Higgs (section~\ref{hsec}),
the bottom (section~\ref{bsec}),  the top (section~\ref{tsec}), and finally
new physics  ---
or better how the lack of new physics is challenging how ideas about naturalness of the Higgs mass (section~\ref{natsec}).

\section{Neutrinos}\label{nu}

\pagina{$\nu_e\leftrightarrow \nu_\mu\leftrightarrow \nu_\tau$ oscillations}\noindent

\begin{figure}
\begin{minipage}{0.45\linewidth}
$$\includegraphics[height=6.4cm]{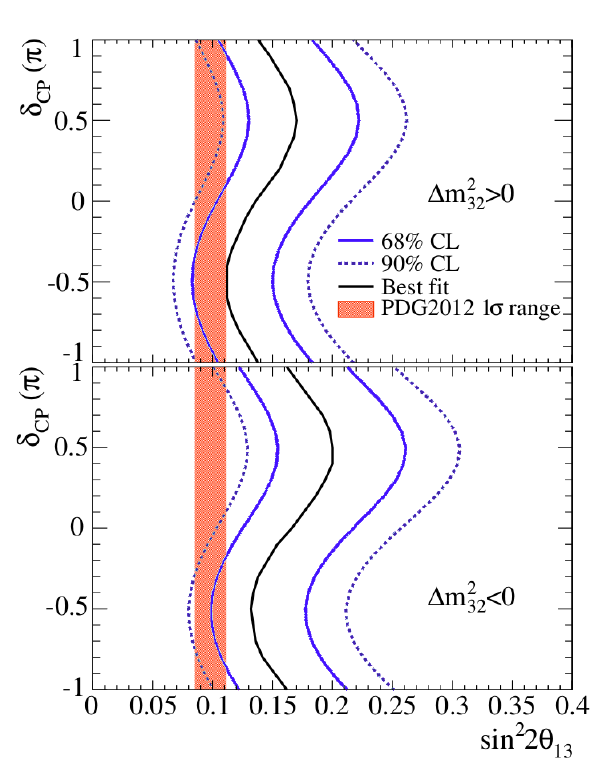}$$
 \caption{Measurements of $\theta_{13}$ from reactors (red vertical band) and from T2K (black/blue bands).\label{fig:q13}}
 \end{minipage}\hfill
 \begin{minipage}{0.45\linewidth}
$$\includegraphics[height=6cm]{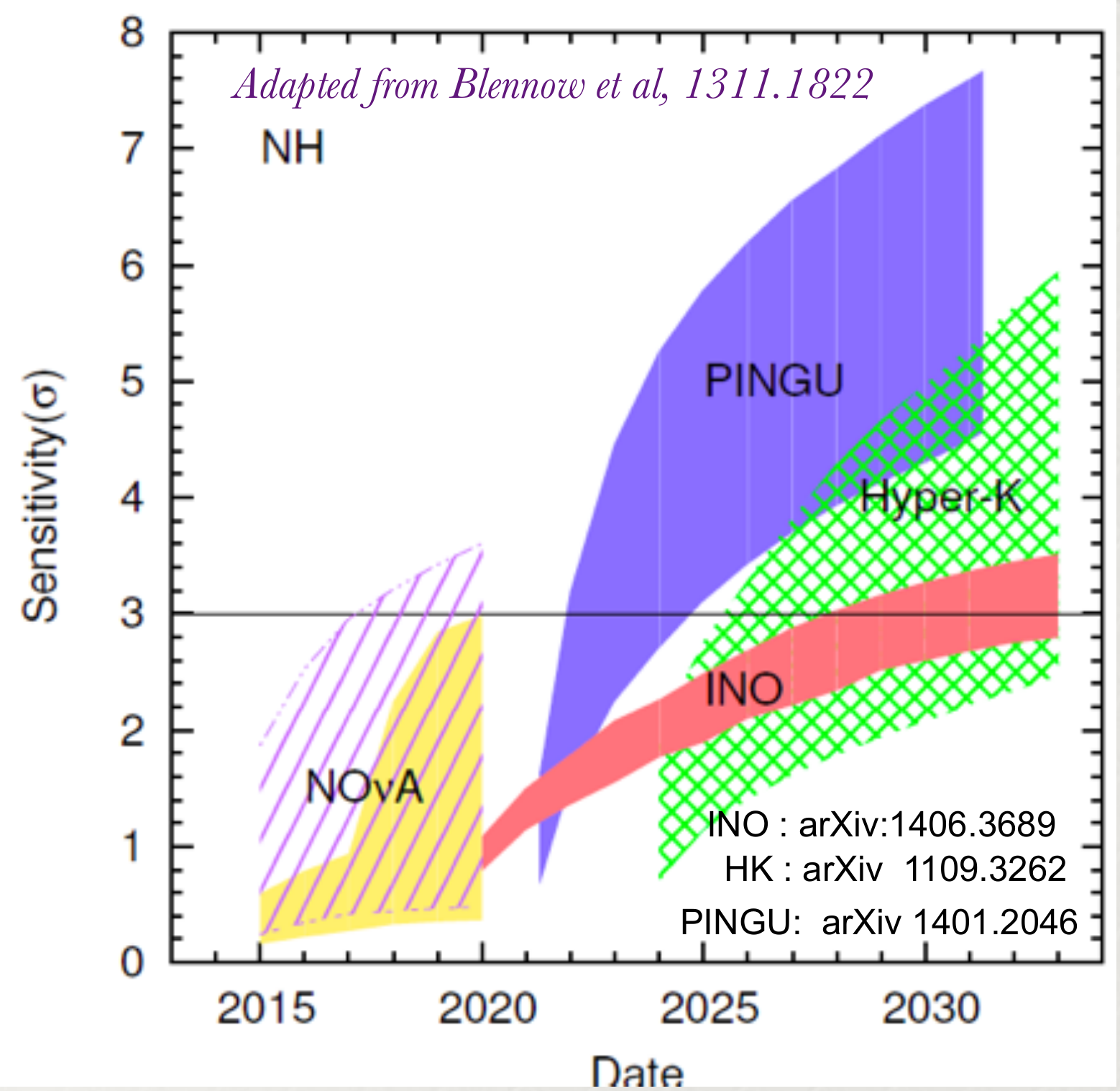}$$
\caption{Future prospects for determining the neutrino mass hierarchy: normal or inverted?\label{num}}
 \end{minipage}\end{figure}

\href{https://indico.in2p3.fr/event/10819/session/0/contribution/45/material/slides/0.pdf}{E. Lisi}$\,$\cite{this} reviewed how a  lot of data are non-trivially explained by $3\nu$ oscillations.
This fascinating field started as astro-physics
and so far gave us 5 new fundamental parameters:
\beq \begin{array}{c}
\Delta m^2_{12} = 7.50\pm 0.20 ~10^{-5}\eV^2,\qquad
|\Delta m^2_{23}| = 2.45\pm0.10~10^{-3}\eV^2 \\[1ex]
\tan^2\theta_{12}=0.452\pm0.035,\qquad
\sin^2 2\theta_{13}=0.087\pm 0.0045 ,\qquad 
\sin^22\theta_{23}=1.02\pm 0.04
\end{array}\eeq
Some years ago, Lisi and others claimed that global fits of solar and atmospheric data favoured at $\sim 1.5\sigma$ level
a non vanishing $\theta_{13}$:  
\beq
 \sin^2 2\theta_{13} = \left\{ \begin{array}{ll} 
 0.05\pm 0.05 & \hbox{CHOOZ and atmospheric~}\\
0.07\pm 0.04 &  \hbox{{\sc KamLAND} and solar}\\  
0.05\pm 0.03 & \hbox{$\nu_\mu\to\nu_e$ at MINOS,  normal hierarchy}\\
0.084 \pm0.005 &  \hbox{$\bar\nu_e\to\bar\nu_e$ at  {\sc DayaBay}}\\
0.113 \pm 0.023 &  \hbox{$\bar\nu_e\to\bar\nu_e$ at  RENO}\\
0.090\pm0.032 & \hbox{$\bar\nu_e\to\bar\nu_e$ at  {\sc Double CHOOZ}}\\
0.140\pm 0.035& \hbox{$\nu_\mu\to\nu_e$ at T2K for $\delta_{\rm CP}=0$, normal hierarchy.}\\
\end{array}\right.\eeq
In 2012 the reactor experiments {\sc DayaBay} and others
confirmed this hint, finding a clear evidence for $\theta_{13}$.
With the most recent data we start to have an indication for a 6th parameter: a non-vanishing CP-violating phase $\delta_{\rm CP}$.
Indeed, as shown in fig.~\ref{fig:q13}, $\theta_{13}$ is measured by reactor experiments,
and a combination of $\theta_{13}$ and of $\delta_{\rm CP}$ is measured by the T2K long-baseline experiment.
The two determinations are in better agreement if
\beq \delta_{\rm CP}\sim -\pi/2. \eeq
So far all observed CP-violating phenomena have the same source: the CKM phase.
The observation of CP violation in neutrino oscillations would double the observed CP-violating parameters.
We have no idea of why CP is violated in nature.
Do complex phases appear everywhere with order one size?

\bigskip

Furthermore, the sign of $\Delta m^2_{23}$ is unknown.
\href{https://indico.in2p3.fr/event/10819/session/0/contribution/19/material/slides/0.pdf}{S. Choubey}$\,$\cite{this}
reviewed how the neutrino mass hierarchy can be determined with
\begin{itemize}
\item[a)]  matter effects in future long-baseline experiments;
\item[b)]  matter effects in future atmospheric $\nu$ experiments; 
\item[c)]  interference between solar and atmospheric oscillations in future dedicated reactor experiments.
\end{itemize}
Future prospects are summarised in figure~\ref{num}.

\pagina{(What) are we learning?}\noindent
Observing a $3\times 3$ chessboard might  be not enough to reconstruct the laws that rule chess.
We observed $3+3$ mixing angles and $3+3+3+2$ fermion masses. With this, we fail to identify the laws that rule flavour.
Unlike in the case of Balmer lines, data show no clean pattern.

Nevertheless, the observed semi-ordered flavour pattern somehow
resembles SU(5) unification with some flavour symmetry acting only on 10$_i$ fermions, such that
\beq \lambda_D\sim\lambda_E \propto \epsilon,\qquad
\lambda_U \propto \epsilon^2,\qquad
\lambda_N \propto\epsilon^0,\qquad
\theta_{\rm CKM} \sim \epsilon,\qquad
\theta_{\rm MNS}\sim 1.
\eeq
But this does not give rise to any sharp prediction.
Only low-level theories (such as texture zeroes or numerology) give precise predictions.
It is maybe interesting to point out that 
$$V_{\rm PMNS}  = R_{12}(\theta_C) \cdot R_{23}(\pi/4) \cdot R_{12}(\pi/4)$$
(where $\theta_C$ is either the Cabibbo angle or the Golden angle $ \theta_{\rm Golden} \equiv {\rm arc cot}\,\varphi^3$)
predicts  neutrino mixings
\beq \theta_{23} = 44.2^\circ,\qquad
\theta_{12}=35.5^\circ\qquad
\theta_{13}=9.3^\circ  \eeq
in agreement with measurments.
When searching if this was already known,  I first found chinese authors  who
studied a similar ansatz, but with $R_{12}(\theta_C) $ on the wrong side.
Later I found that other chinese authors put $R_{12}(\theta_C) $ on both sides, obtaining another wrong prediction.
Finally, I found that more chinese authors who basically got the same prediction~$\,$\cite{Zheng}.

\begin{figure}[t]
$$\includegraphics[height=5.3cm]{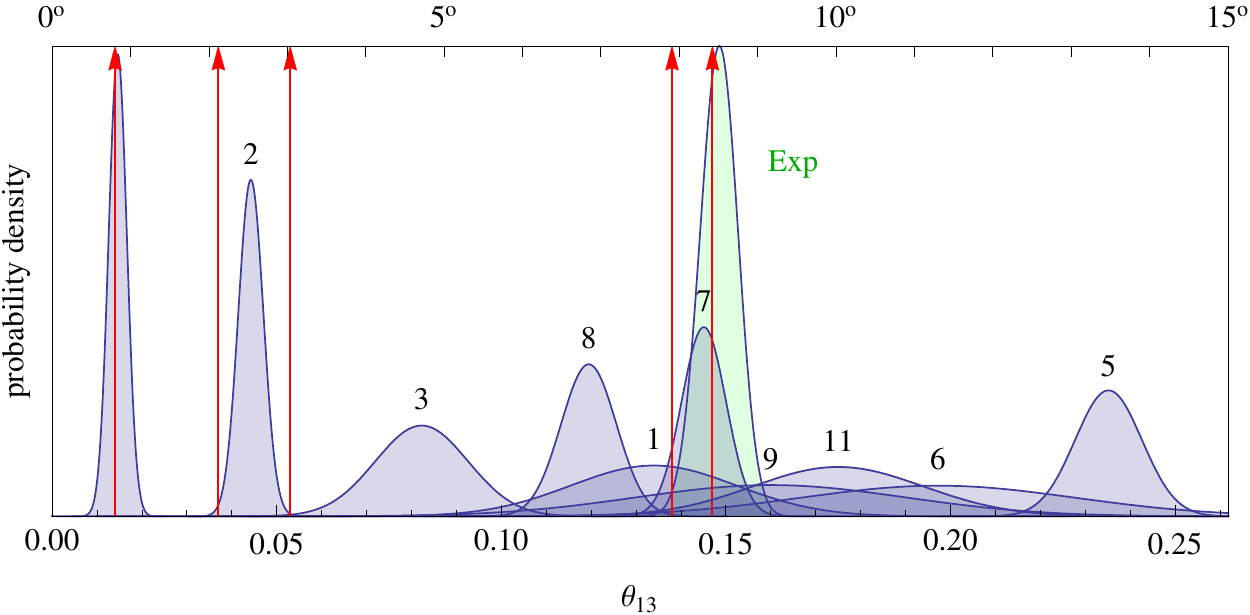}$$ 
\caption[]{Compilation of predictions for $\theta_{13}$: the  bells show the central value and (when possible) the uncertainty of
each prediction.  The green bell is the observed value of $\theta_{13}$.\label{q13}}
\end{figure}

What is the lesson?
Fig.~\ref{q13} shows a compilation of predictions for $\theta_{13}$:
there are so many of them that any measured value would have agreed with some prediction.

 \pagina{The rector $\bar\nu_e$ anomaly}
A re-computation of   $\bar\nu_e$ fluxes emitted by fission reactors
claims that measured rates are $\sim5\%$  lower than predicted unoscillated rates
and that the theoretical uncertainty is at the $1-2\%$ level.

The deficit could be due oscillations into extra sterile neutrinos $\nu_s$.
\href{https://indico.in2p3.fr/event/10819/session/2/contribution/29/material/slides/0.pdf}{S. Zsoldos}$\,$\cite{this} presented STEREO:
a new  reactor experiment with short base-line $L\sim 10\,{\rm m}$
to search for sterile oscillations with $\Delta m^2 \sim \eV^2$.
Indeed, present null results  push $m_{\nu_s}\circa{>}\eV$.
However, such a large sterile neutrino mass is already disfavoured by standard cosmology, as discussed by \href{https://indico.in2p3.fr/event/10819/session/1/contribution/48/material/slides/0.pdf}{N. Saviano}$\,$\cite{this}.

Furthermore, 
 the measured $\bar\nu_e$ reactor spectra differ also from the predicted spectra at the same level,
 showing a `bump'  that cannot be due to oscillations.
  \href{https://indico.in2p3.fr/event/10819/session/0/contribution/74/material/slides/0.pdf}{A. Hayes}$\,$\cite{this} reviewed the issue:
nuclear theory and data seem to be not solid enough to allow the claim that there is an anomaly.
Modern reactor experiments bypass these uncertainties by observing oscillations as
differences between the energy spectra measured in a near detector and a far detector. 
 
\pagina{Neutrino-less double-beta decay}\noindent
If neutrino masses are of Majorana type, neutrino-less double-beta decay
is the only realistic way of observing lepton-number violation in the electron sector:
the amplitude is proportional to  $m_{ee}$, the $ee$ entry of the Majorana neutrino mass matrix.

Oscillation experiments imply a detectable $0\nu2\beta$ rate if neutrinos have inverted mass hierarchy.
However the computation of $0\nu2\beta$ rates involves difficult and uncertain nuclear physics.
\href{https://indico.in2p3.fr/event/10819/session/0/contribution/36/material/slides/0.pdf}{P.K. Raina}$\,$\cite{this} reviewed 
the issue of nuclear matrix elements. The  $0\nu2\beta$ rate is computed as
\beq \Gamma_{0\nu2\beta} =G~\left|\mathscr{M}   m_{ee}/m_e  \right|^2\eeq
where $G$ is known phase space factor and
$\mathscr{M}$ is the nuclear matrix element.
We heard that ``NTEM  QRPA, PHFB, DHF, EDF, IBM, SSDH, OEM'', which,  in the language of
nuclear physics, means ``there are various approaches, but no agreement''.
Even worse, some experts think that 
 the axial nucleon coupling $g_A$,
measured in vacuum to be $g_A\approx 1.25$,
can acquire inside nuclei a smaller value, $g_A\approx 1.25 A^{-0.15}$.
In the worst case,
such nuclear quenching of $g_A$ could reduce $0\nu2\beta$ rates by up to 6 $-$ 50,
making all experiments much less sensitive than what initially thought$\,$\cite{Iachello}.

\begin{table}
$$\tiny
\hspace{-0mm}\begin{array}{c|ccccccc}
\hbox{nucleus} &
^{76}{\rm Ge} &
^{82}{\rm Se}&
^{100}{\rm Mo} &
^{130}{\rm Te} &
^{136}{\rm Xe}&
^{150}{\rm Nd}\\ \hline
\hbox{$Q$ in keV} &  2039 &  2995 & 3034  &  2529&2476&3367 \\
\hbox{$T_{1/2}^{2\nu2\beta}$ in $10^{20}$ yr} & 15\pm1 & 0.92\pm0.07 & .071\pm.004 & 9\pm1 &23\pm1 & .082\pm .009\\[2mm]
\hbox{$G$ in $10^{-14}/$yr} & 0.63 & 2.73 & 11.3 & 4.14 &  4.37 & 19.4\\  \hline
\mathscr{M} \hbox{~[IBM-2 2015]} & 4.68 & 4.37 & 3.7  & 4.03 & 3.33& 2.32 &\\
\mathscr{M} \hbox{[\v{S}FRVE 2007]} & 3.3\div 5.7 & 2.8\div 5.1 & 2.2 \div 4.6 & 2.3 \div 4.3& 1.2\div 2.8 &\\
\mathscr{M} \hbox{~~~~~~[CS 2009]} & 4.0\div 6.6 & 2.8\div 4.6 & 2.7 \div 4.8 & 3.0\div 5.4 &2.1 \div 3.7 & \\
\mathscr{M} \hbox{~[MPCN 2008]} & 2.3 & 2.2 & & 2.1 & 1.8 &\\
\mathscr{M} \hbox{~~~~~~~[BI 2009]} & 5.5 & 4.4 & 3.7 & 4.1 &&2.3 \\
\mathscr{M}_0 \hbox{~~~[SMK 1990]}&4.2 & 4.0 & 1.3 & 3.6 & 1.7 &0.6\\ \hline
\hbox{90\%CL bound on} & 190, 160, 210 &  4.4 &  5.8 & 30 & 260 & 0.036\\
\hbox{$T_{1/2}^{0\nu2\beta}$ in $10^{23}$ yr} &\hbox{ HM, IGEX,GERDA}\!\! &\hbox{ NEMO-3} & \hbox{ NEMO-3}  &\!\! \hbox{\sc Cuoricino} \!\! & \hbox{ KL-Zen, EXO}&\hbox{ NEMO-3} \\[2mm]
\hbox{\color{blue}Bound on $m_{ee}/h$} &\color{blue} 0.35, 0.38, 0.33\eV &\color{blue}1.2 \eV& \color{blue}1.5 \eV&\color{blue}0.40\eV & \color{blue}0.29 \eV&\color{blue}31\eV\\
\end{array}$$
\caption{\label{0nu}Summary of $0\nu2\beta$ parameters and experimental data and bounds.}
\end{table}

\color{black}

\subsection{High-energy neutrinos}
 
 \href{https://indico.in2p3.fr/event/10819/session/1/contribution/61/material/slides/0.pdf}{A. Vincent}$\,$\cite{this} discussed flavour reconstruction 
 of neutrino-induced events observed at {\sc IceCube}:
after taking into account track mis-identification, all oscillated neutrino production mechanisms 
($\pi$ decays, $\mu$ decays, $n$ decays) can fit the data.

 \href{https://indico.in2p3.fr/event/10819/session/1/contribution/15/material/slides/0.pdf}{M. Masip}$\,$\cite{this}
argued that the excess over the neutrino atmospheric background observed by {\sc IceCube} seems to be
showers, rather than tracks.
In theoretical language, this means Neutral Current events, rather than Charged Current events. 
Consequently he proposed that the claimed excess 
can be fitted by new physics that increases the NC cross sections of ordinary atmospheric neutrinos;
in particular models with extra dimensions predict double-bang events.

 \href{https://indico.in2p3.fr/event/10819/session/0/contribution/38/material/slides/0.pdf}{J. Heeck}$\,$\cite{this} reviewed gauged U(1)$_{B-L}$.  If
broken by 4 units, by an $\nu_R^4$ operator,  one would obtain
$pp\to 4\ell$ signals; 
Dirac leptogenesis would be allowed, giving an excess of neutrinos in the Cosmic Microwave Background, $N_{\rm eff}\approx 3.14$.
\href{https://indico.in2p3.fr/event/10819/session/0/contribution/24/material/slides/0.pdf}{J. Lopez Pavon}$\,$\cite{this} discussed  type-I see-saw
in the full range of possible right-handed neutrino masses, finding that
$1\eV < m_{\nu_R}<100 \MeV$ is excluded by cosmology because $\nu_R$ would thermalise.
\href{https://indico.in2p3.fr/event/10819/session/1/contribution/20/material/slides/0.pdf}{V. de Romeri}$\,$\cite{this} 
discussed type-I see-saw signal at a future circular $e^+e^-$ collider,
finding that a $Z\to\mu\tau$ rate could be produced  at an observable level,
while an observable $Z\to e\mu$  is already excluded by $\mu\to e\gamma$ and related measurements.


%
%


\section{Dark Matter} \label{DMsec}
In the past century we understood matter.  Cosmology and astrophysics tell that
there is more matter left to be understood,  called Dark Matter.

For some time `neutralino' was used as a synonymous of `Dark Matter',
but  theoretical progress showed that the DM mass can be in the wider range
$M_{\rm DM} = 100\GeV \times 10^{\pm 40}$.
This means that DM could be macroscopical objects to be studied as astrophysics
(although cosmology disfavours this possibility),
DM could be an unobservable Planck-scale particle
(to be studied by string theorists);
DM could be a weak scale  particle
(to be studied by particle physicists),
DM could be lighter 
(in the energy range of nuclear physics down to chemistry),
DM could be an axion-like ultra-light scalar 
(to be studied as classical field theory),
demanding a variety of experimental approaches.

Let us summarise what is known.

So far we  observed Dark Matter gravity. 
We only have upper bounds on any other DM interaction with matter and among itself.
In cosmology, this lack of interactions is what allows DM to cluster forming galaxies,
 while normal matter is still interacting with non-clustering photons.
Cosmological data are reproduced if DM has a primordial inhomogeneity of adiabatic type:
a single dominant primordial perturbation equal for all components of the universe.
Loosely speaking, this suggests some connection with matter.
The assumption that DM is a thermal relic favours
$M_{\rm DM}   \circa{<}100\TeV$: then DM could be discovered in  direct, indirect, collider
searches around 2010 or maybe later.
We passed the golden moment where {\sc Pamela, Fermi, AMS, CDMS, Xenon, LHC} started to produce data: backgrounds are starting to become a limitation.

\pagina{Dark Matter searches are reaching the backgrounds}\noindent

\begin{figure}
$$\includegraphics[width=0.65\textwidth]{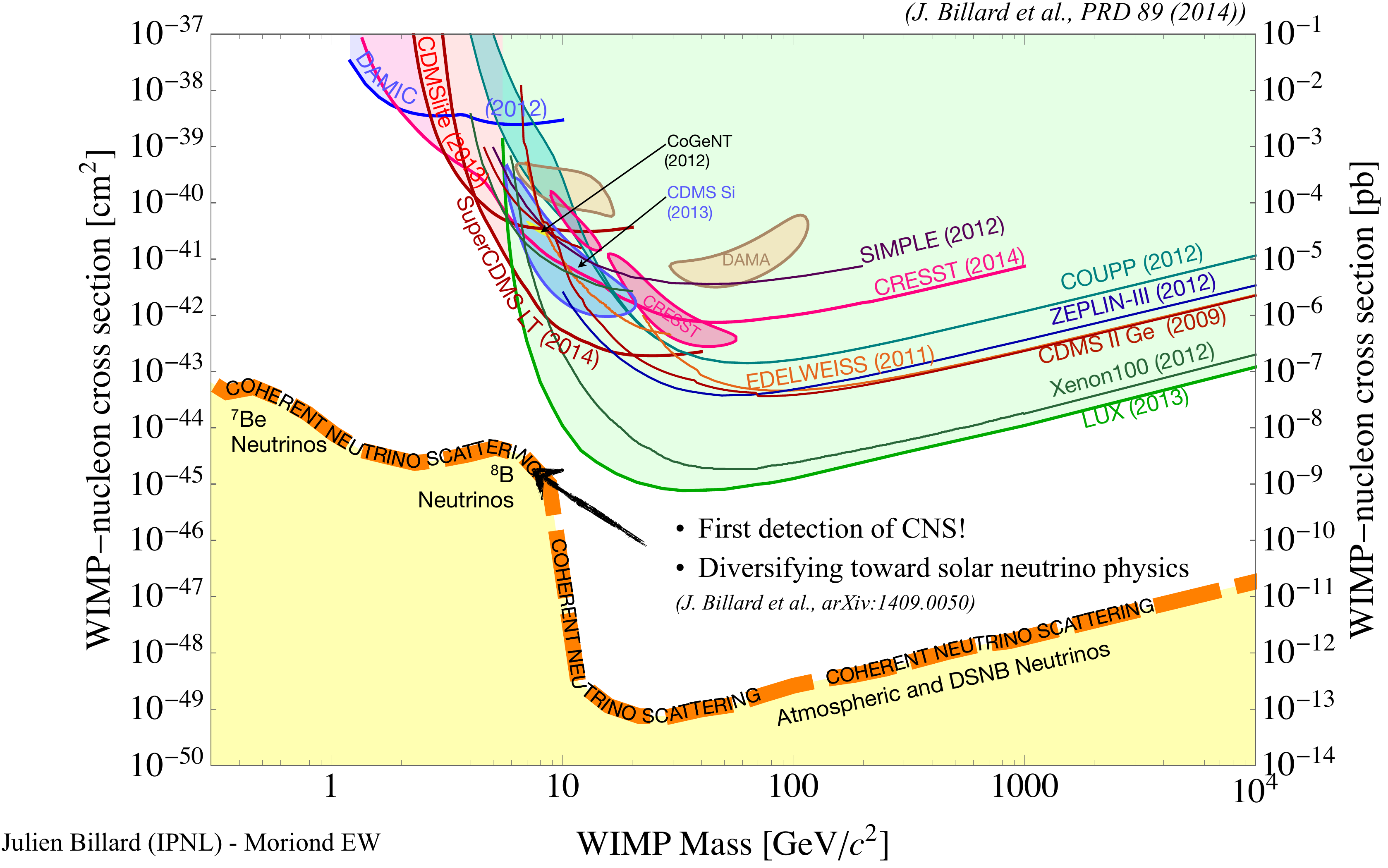}$$
\caption{\label{DM}  The neutrino background to direct DM searches.}
\end{figure}

{\em Direct searches} for DM with electro-weak mass improved by 3 orders of magnitude in the past decade and
are now 3-4 orders of magnitude above the irreducible  $\nu$ background,
as shown in fig.~\ref{DM}.
This was discussed by \href{https://indico.in2p3.fr/event/10819/session/1/contribution/33/material/slides/0.pdf}{J. Billard}$\,$\cite{this} who proposed
how, in case of no signal above the background, we might survive to the crash on the ground, relying on
seasonal variations (DM rates are maximal in June, while solar $\nu$ rates are maximal in January), multiple experiments,
directionality of scattered particles.
Various authors explore how to improve searches for sub-GeV DM particles:
present bounds are still far from the irreducible $\nu$ backgrounds.

\medskip

{\em Indirect searches} aim at detecting in cosmic rays an excess of
$e^+$, $\bar p$, $\gamma$ possibly produced by DM annihilations or decays.
In the past years such cosmic rays have been observed at weak scale energies,
and no undisputable evidence of a DM excess above the astrophysical fluxes was found.
The uncertain uncertainties on such backgrounds to DM searches are now the limiting factor:
it will be difficult to clarify the tentative claims and to improve the searches.
Once all data will be available, it will maybe possible to better understand galactic astro-particle physics,
improving DM searches.


%

\medskip


{\em Searches at colliders} mostly rely on the classical  DM signal 
\beq\label{DMs} \hbox{missing energy + something to tag the event  (a jet, a $\gamma$, ...).}\eeq
At a hadron collider,
this signal can be easily seen if DM is produced with QCD-like cross section, like in supersymmetric models.
On the other hand, the same signal can be easily  missed below the backgrounds,
if instead the cross section for DM production is (loosely speaking) below the electro-weak neutrino cross section.

Furthermore, DM can easily be too heavy for LHC.  This is for example the case of
Minimal DM models, where DM only has  electro-weak gauge interactions:
assuming that the cosmological DM abundance is reproduced as thermal relic,
one predicts a TeV-scale DM mass.

\medskip

The TeVatron and LHC experimental collaborations searched for the signals in eq.~(\ref{DMs}) without finding new physics.
Various  bounds have been presented in the context of
effective contact operators between DM and quarks,
suppressed by a scale $\Lambda$:
this is appreciated as a general framework that allows
collider experimentalists to present results in the same ($M_{\rm DM}, \sigma_{\rm SI/SD}$)
plane used in direct DM searches.
However:

\begin{enumerate}
\item {The effective operator approximation fails at LHC.}

Effective operators are a valid approximation at energies below $\Lambda$.
For any collider the sensitivity to $\Lambda$ will be much below the collider energy,
because tagging the invisible
signal needs extra $j$ or $\gamma$, implying a small cross section.


\item  {Effective operators could mislead to miss the missing energy signal.}

The assumed growth of $\sigma \sim E^2/\Lambda^4$ with energy
is crucial in making high-energy colliders competitive with direct searches.
But in models where $1/\Lambda^2 \approx g_{\rm mediator}^2/M_{\rm mediator}^2$,
such growth stops at the mediator mass:
the signal is no longer at the highest energy.

\item What LHC would first see is the heavy mediator particle,
not missing energy.   

Even using ``simplified models'' the casistics of possible mediators is tedious.  The basic possibilities are
a colored mediator in $t$-channel (that would be pair produced with a large QCD cross section)
or a neutral mediator in $s$-channel (that would also give a peak in the $pp\to jj$ cross section).
In both cases the mediator gives much cleaner signals than Dark Matter. 

\end{enumerate}
This was described by many recent works, and there is presently interest in systematizing  
``simplified models'' that capture a minimal and sufficient amount of new physics.
This approach was described by
\href{https://indico.in2p3.fr/event/10819/session/9/contribution/114/material/slides/0.ppt}{B. Zaldivar}$\,$\cite{this}.
\href{https://indico.in2p3.fr/event/10819/session/1/contribution/88/material/slides/0.pdf}{U. Hash}$\,$\cite{this} emphasised how higher order corrections for DM search at LHC can be relevant.

\pagina{Models for DM}\noindent
Building models for DM that satisfy the present theoretical and experimental restrictions
is not particularly difficult.
At the Moriond 2015 conference
\href{https://indico.in2p3.fr/event/10819/session/0/contribution/0/material/slides/0.pdf}{A. Ahrice}$\,$\cite{this} and 
\href{https://indico.in2p3.fr/event/10819/session/2/contribution/22/material/slides/0.pdf}{D.R. Lamprea}$\,$\cite{this} presented
models that simultaneously can account for observed DM and for observed neutrino mass.

\smallskip

Furthermore,
 \href{https://indico.in2p3.fr/event/10819/session/1/contribution/32/material/slides/0.pdf}{G. Arcadi}$\,$\cite{this} proposed a model of decaying DM, containing the Yukawa couplings
$$\hbox{$y$ matter DM scalar + $y'$ matter matter scalar}.$$
$y$ gets predicted by assuming that the cosmological DM abundance is reproduced through the freeze-in mechanism,
implying a scalar long lived compared to collider time-scales.


\section{The Higgs} \label{hsec}

 \href{https://indico.in2p3.fr/event/10819/session/5/contribution/5/material/slides/0.pdf}{R. Kogler}$\,$\cite{this} summarized global fits
of electroweak precision data.  The summary of the summary is: `{SM ok}'.

\begin{figure}
$$\includegraphics[width=0.47\textwidth]{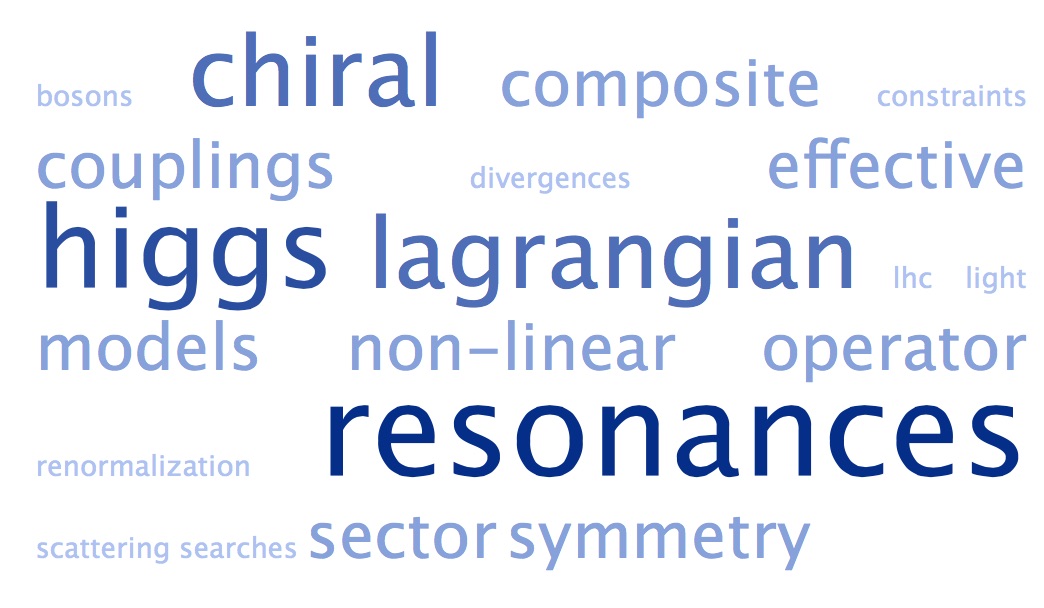}$$
\caption[]{Theoretical summary of talks about Composite Higgs.\label{cloud}
\href{https://indico.in2p3.fr/event/10819/session/8/contribution/17/material/slides/0.pdf}{D. Espriu}$\,$\cite{this} discussed how unitarity can constrain resonances.
\href{https://indico.in2p3.fr/event/10819/session/4/contribution/23/material/slides/0.pdf}{K. Kanshin}$\,$\cite{this} discussed operators
with arbitrarily high dimensions characteristic of non-linearities 
assuming a chiral Lagrangian for the Higgs.
\href{https://indico.in2p3.fr/event/10819/session/4/contribution/50/material/slides/0.pdf}{I. Brivio}$\,$\cite{this} discussed typical effects of such nonlinearities.
\href{https://indico.in2p3.fr/event/10819/session/5/contribution/78/material/slides/0.pdf}{A. Kaminska}$\,$\cite{this} considers techni-$\rho$ in (partially)composite Higgs models.
\href{https://indico.in2p3.fr/event/10819/session/3/contribution/104/material/slides/0.pdf}{F. Riva}$\,$\cite{this} reviewed  
new physics signals as dimension 6 operators: there are 79 of them  (2499 with flavour).
Given present experiments,
NRO are mostly equivalent to simpler $\kappa$ modifications of $h$ and $Z$ SM couplings.}
\end{figure}

Various authors presented general aspects of composite Higgs models.
Since I never worked on this subject, the  theoretical summary in fig.~\ref{cloud} 
has been obtained by processing  the slides with a computer code that identifies the key words.
Below, I would like to explain my doubts about the field of composite Higgs models.

%

\pagina{Is the Higgs  composite?}\noindent
Now that the SM scalar boson was  found and that its properties started to be measured,
this question is obviously a hot topic for experimentalists.
\href{https://indico.in2p3.fr/event/10819/session/3/contribution/69/material/slides/0.pdf}{A. David}$\,$\cite{this}
emphasised that future data will allow to go beyond overall rates,
where new physics effects are  parameterized by overall $\kappa$ factors.
Already at LHC run I experimentalists measured angular correlations and transverse momentum distributions that confirm that 
the resonance at 125 GeV is really the scalar Higgs, rather than some impostor that fakes the same total rates.
Given the possibility that some new physics might cancel out  in the total rate,
it would be useful to identify extra useful  pseudo-observables, that describe how the rates depend
on kinematics.


\medskip

Since decades, naturalness motivates a significant theoretical activity on composite Higgs models of a special type:
models where the Higgs is {\it composite at the weak scale and made of fermions}.

However
LEP precisely measured  3 components of the Higgs doublet $H$
(those eaten by the $Z$ and by the $W^\pm$), setting a strong bound
$\Lambda\circa{>}10\TeV$
on the compositeness scale
 that suppresses the $|H^\dagger D_\mu H|^2$
 effective operator that violates the custodial symmetry, accidentally present in the SM.
LHC finally discovered the last component of $H$, the physical Higgs $h$,
allowing to set weaker constraints on a wider set of effective operators.
As a consequence composite Higgs models need some fine-tuning to achive electroweak-symmetry breaking
compatibly with data.


More importantly, the Higgs does much more than breaking the electroweak symmetry.
The SM elementary Higgs also has a set of Yukawa couplings that non-trivially reproduce the observed
 pattern of lepton and quark masses and of flavour-violating processes.
It looks impossible to find a theory at the weak scale that dynamically generates all this flavor structure,
while at the same time giving no new physics effects in flavour observables.
 
\medskip
 
In this situation,  theory of composite Higgs degenerated in `cosettology'.
While the $\SU(3)_L\otimes\SU(3)_R\to\SU(3)_V$ dynamical symmetry breaking pattern of QCD
is understood as the result of having 3 light quarks, 
composite Higgs theories postulate ad-hoc properties 
of  the unknown extra strong dynamics  in order to 
recover the custodial symmetry (this looks plausible), to 
satisfy flavour bounds (partial compositeness is invoked), etc.
All of this gets described in a formal way using general techniques (chiral effective Lagrangians, AdS/CFT, approximations to QCD...).
In absence of a fundamental theory with the structure needed to accomodate all null results, it is not clear 
how much the deep words in fig.~\ref{cloud} really go beyond a purely phenomenological approach, useful when  data are coming.

As far as I know, one could make real  theories of composite Higgs by giving up on naturalness:
\begin{itemize}
\item[a)]  One possibility is accepting a very large compositeness scale (for example $10^{10}\GeV$ such that axion-Higgs unification
becomes possible).

\item[b)]  Another possibility is making TeV-scale theories using elementary {techni-scalars} ${\cal H}$
and techni-fermions ${\cal Q}$ with techni-Yukawas
$y_{ij} \, \psi_i {\cal H}_j {\cal Q}$ to SM fermons $\psi$, realising partial compositeness.
 \end{itemize}

\pagina{More Higgs}

 \href{https://indico.in2p3.fr/event/10819/session/3/contribution/51/material/slides/1.pdf}{A. Celis}$\,$\cite{this} studied models with 2 Higgs doublets and  Yukawa matrices proportional to each other (at least at tree level) in order to avoid unobserved flavor-violating effects.
 \href{https://indico.in2p3.fr/event/10819/session/8/contribution/89/material/slides/0.pdf}{S.I. Godunov}$\,$\cite{this}
 considered Higgs triplets present in type-II see-saw models, 
 discussing signals where the triplet decays into $hh$;
furthermore he pointed out how  higher rates can be obtained with an ad-hoc fine tuning proposed by
Georgi and Machacek.
  \href{https://indico.in2p3.fr/event/10819/session/4/contribution/6/material/slides/0.pdf}{V. Bizouard}$\,$\cite{this} performed
  one-loop computation of the system of 3 scalar Higgses present in the NMSSM.
This was the first talk about supersymmetry.

\medskip

\href{https://indico.in2p3.fr/event/10819/session/5/contribution/100/material/slides/0.pdf}{M. Neubert}$\,$\cite{this}  presented predictions for
rare $Z,W$ decays into exclusive channels, that 
 can be precisely computed using factorisation 
and measured with high luminosity LHC:
$$\includegraphics[width=0.8\textwidth]{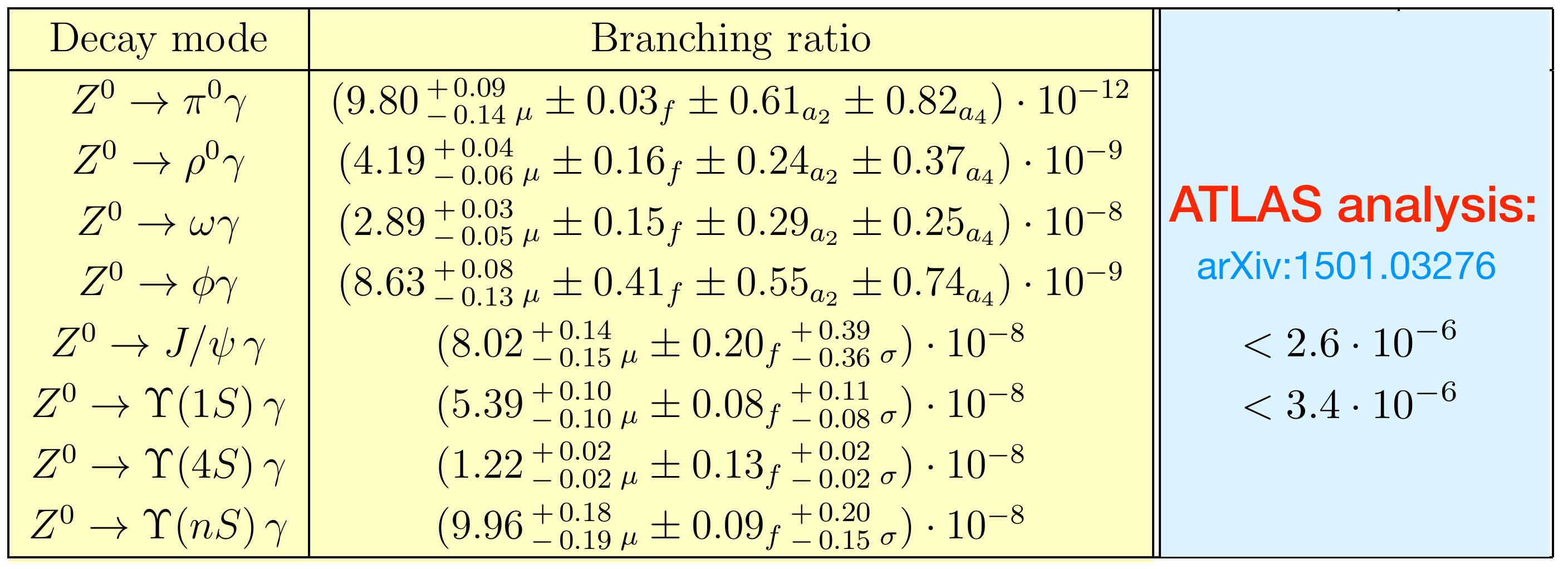}$$
Furthermore, Neubert discussed how searches for $h\to J/\psi~\gamma$ and $h\to\Phi\gamma$ 
could somehow constrain  the small Higgs couplings to the charm quark and maybe the even smaller coupling to the strange quark.

\section{Bottom quark}\label{bsec}

\begin{figure}
$$\includegraphics[width=0.38\textwidth]{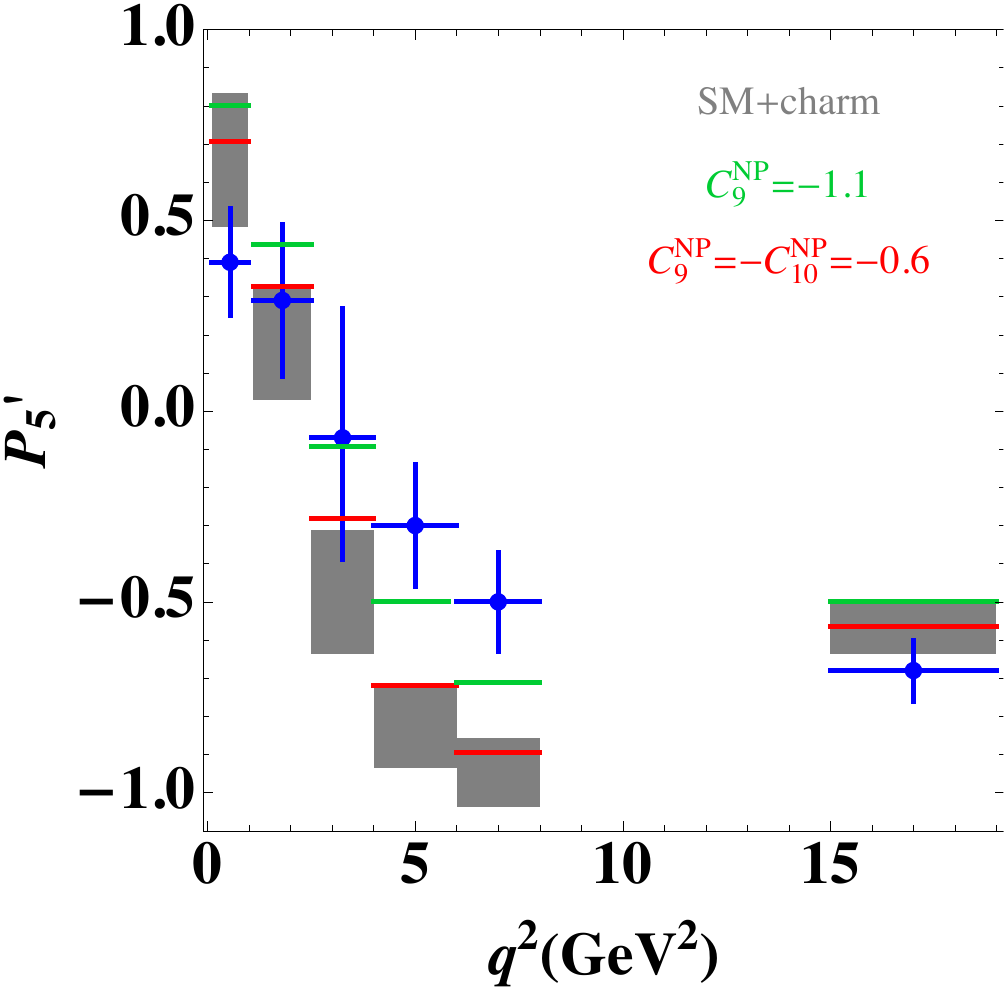}\qquad
\includegraphics[width=0.55\textwidth]{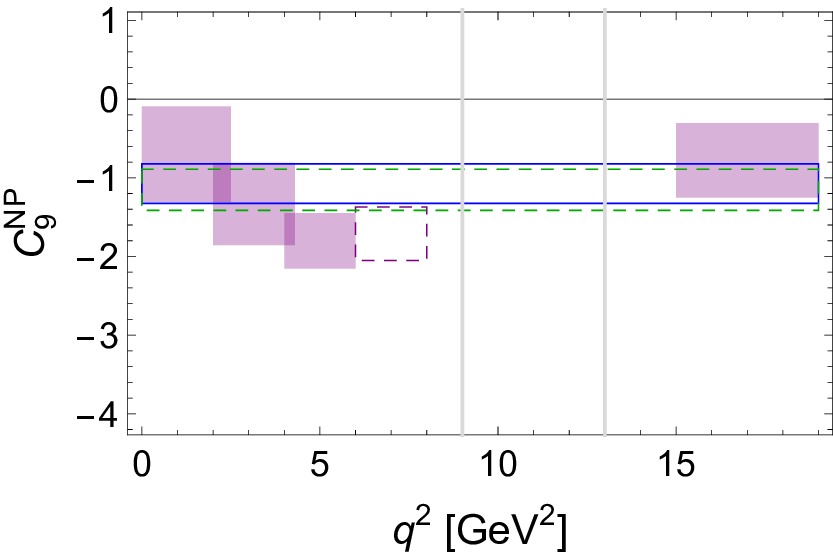}$$
\caption{\label{B} Left: data from the LHC-$B$ collaboration about some angular distribution parameter in $B\to K^*\mu^+\mu^-$ 
compared to SM predictions, and to fits with new physics beyond the SM (thanks to J. Matias).
Right: $q^2$-dependence of the anomaly in $B\to K^*\mu^+\mu^-$ decays (thanks to D. Straub).}
\end{figure}

\pagina{The $B\to K^*\mu^+\mu^-$ anomaly}

This is the topic that attracted most attention, given that 
\href{https://indico.in2p3.fr/event/10819/session/10/contribution/84/material/slides/0.pdf}{the LHC-$B$ collaboration}$\,$\cite{this} 
presented, for the first time,
the full LHC run I data  about angular distributions in $B\to K^* \mu^+\mu^-$ decays.
The anomaly found in the previous partial data release persists: the central values got closer to present SM predictions, with smaller experimental uncertainties.

In order to understand what goes on, one needs to perform a global analyses of this data together with data about related 
$B\to K\ell^+\ell^-$  and other total decay rates.
Luckily,
\href{https://indico.in2p3.fr/event/10819/session/10/contribution/14/material/slides/0.pdf}{J. Matias}$\,$\cite{this,Matias} and
\href{https://indico.in2p3.fr/event/10819/session/10/contribution/87/material/slides/0.pdf}{D. Straub}$\,$\cite{this,Straub} 
obtained the LHC-$B$ data,
disappeared from coffee breaks and ski slopes, and, on the final day of the conference,
reappeared to present the results of their  analyses, 
that favour new physics beyond the Standard Model at 
$\approx 4\sigma$ confidence level.   
This caused more panic than the solar eclipse that took place during their talks.

Their two independent analyses agree on this result, and on the claim that the needed
new physics can be the following effective operator
\beq \frac{ (\bar b \gamma_\mu s)_L (\bar\mu \gamma_\mu \mu)_{L\hbox{ or }V}}{(20-30\TeV)^2}.\label{op}
\eeq
Fig.~\ref{B}a shows the LHC-$B$ data, the SM prediction, and how this operator can improve the agreement with data.
According to the analysis by Matias, reasonable fits are obtained also if the quark current is right-handed. 
\href{https://indico.in2p3.fr/event/10819/session/10/contribution/108/material/slides/0.pdf}{A. Crivellin}$\,$\cite{this} discussed how this kind of operator can be  mediated at tree level by lepto-quarks or by a $Z'$ vector coupled to $L_\mu-L_\tau$.

\smallskip

The big question is if QCD uncertainties, possibly related to some unknown
non-perturbative non-factorizable charm quark loop effect, 
have been under-estimated and could bring the SM prediction in agreement with data.
The charm loop would give a left-handed quark current (due to the coupling to the $W$)
and a vectorial lepton current (due to the coupling to the photon).
This kind of structure is compatible with data, see eq.~(\ref{op}).
Furthermore, fig.~\ref{B}b shows how the anomaly depends on the transferred momentum
$q^2$:
the effective operator of eq.~(\ref{op}) would give a constant; while the charm could give some unknown
$q^2$-dependent effect.  Both possibilities are compatible with data.
Finally, the charm loop would give an effect equal in $\mu^+\mu^-$ and in $e^+e^-$,
while new physics could violate lepton universality e.g.\ in
\beq R_K =\frac{{\rm BR} \left ( B^+ \to K^+ \mu ^+ \mu ^-\right )}{{\rm BR} \left ( B^+ \to K^+ e^+ e^-\right )}= 0.745\pm 0.09_{\rm stat} \pm0.036_{\rm syst}.
\eeq
New physics could even violate lepton flavour giving $B\to K^{(*)} \mu e$ decays.
So far data about rates favour an effect present only in muons;
but from an experimental point of view measuring electrons is more difficult.

In conclusion, it seems that future data can clarify the situation.

\subsection{More flavour}

\href{https://indico.in2p3.fr/event/10819/session/10/contribution/25/material/slides/0.pdf}{M. della Morte}$\,$\cite{this} reviewed lattice
predictions for flavour physics.
\href{https://indico.in2p3.fr/event/10819/session/10/contribution/21/material/slides/0.pdf}{R. Knegjens}$\,$\cite{this} reviewed   the SM
predictions for $K^+\to \pi^+\nu\bar\nu$
and
$K_L \to \pi \nu\bar\nu$, and how these decays (with experimental results coming soon) are golden channels for probing new physics.
\href{https://indico.in2p3.fr/event/10819/session/11/contribution/35/material/slides/0.pdf}{W. Dekens}
reviewed left-right symmetric extensions of the Standard Model and their signals in $B$ and $K$ mixing.

\section{The top quark}\label{tsec}
A precise measurement of the top quark mass
(or better, of the top quark Yukawa coupling $y_t$, which is the fundamental parameter, well defined at loop level even including weak corrections) is important because
 {$y_t$ is the biggest coupling of the Higgs}, relevant for
 naturalness; for stability bounds on the SM Higgs potential;
for MSSM predictions of the Higgs mass...

\smallskip

\begin{figure}
 $$\includegraphics[width=0.56\textwidth]{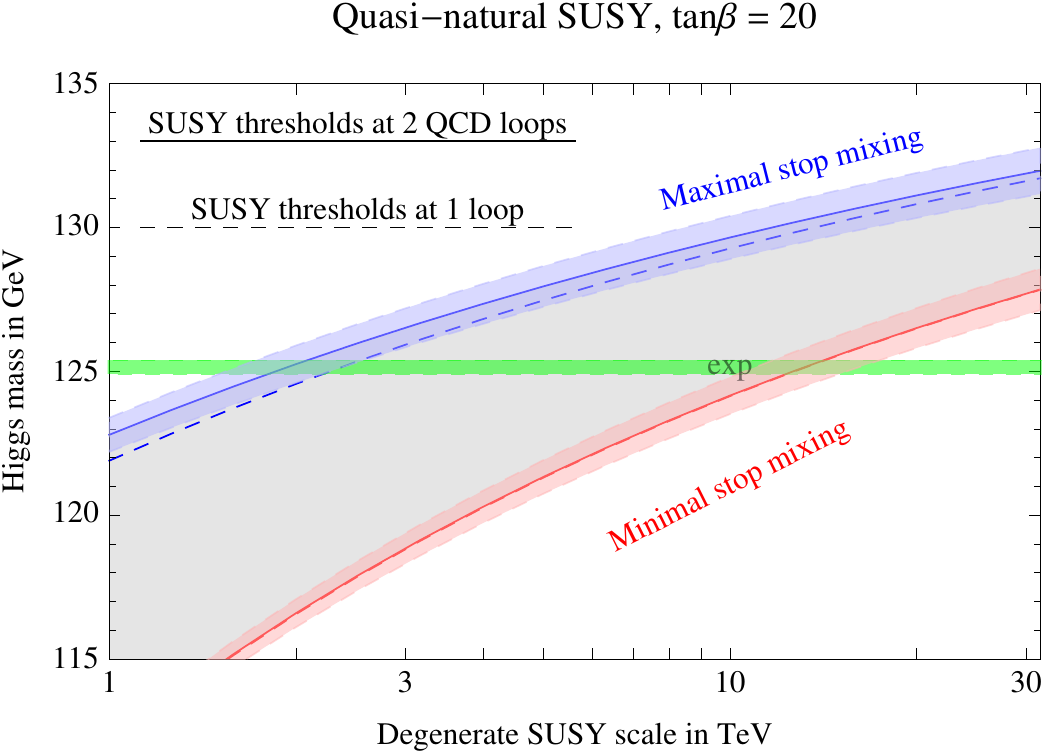}\qquad
 \includegraphics[width=0.39\textwidth]{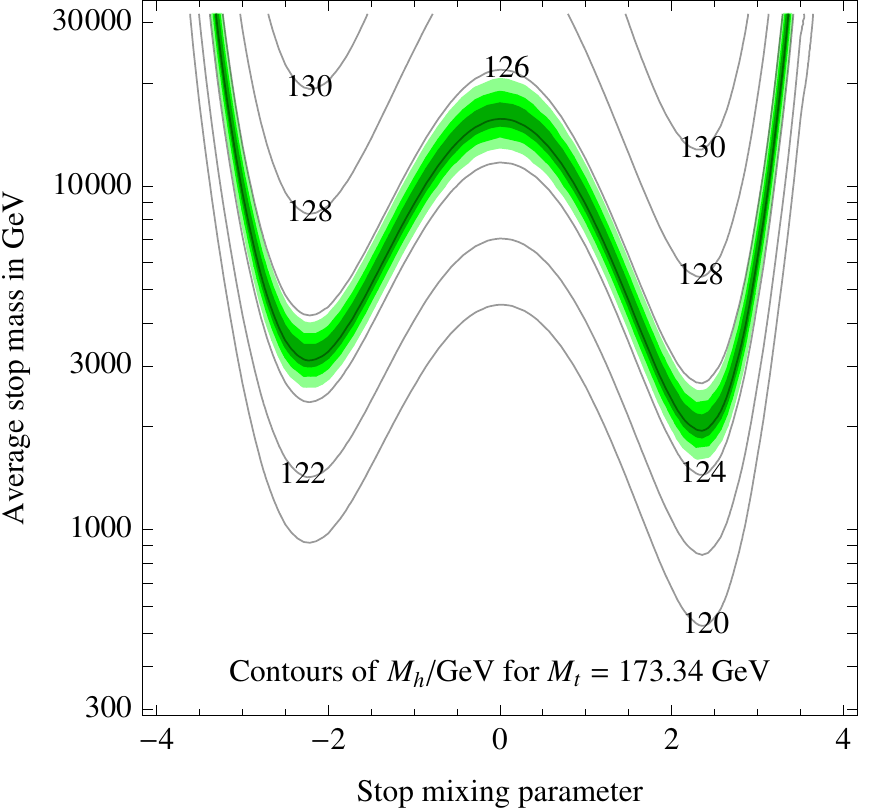}
 $$ 
\caption{\label{susy}Precise MSSM prediction for the Higgs mass in the limit of degenerate sparticles.
}
 \end{figure}

Concerning this latter issue, the use of effective field theory techniques, that apply if the SUSY scales is above about 1 TeV,
allowed to decrease the theoretical uncertainty on $M_h$ from $\pm3 \GeV$ down to about $\pm 1\GeV$.
The predicted $M_h$ is a few GeV below the output of  codes based on the diagrammatical approach:
consequently the SUSY scale needed to reproduce the observed $M_h$ becomes heavier and is in the multi-TeV region, as shown in fig.~\ref{susy}.$\,$\cite{Mh}
The dominant uncertainty on the $M_h$ prediction is now the uncertainty on $M_t$.

\smallskip

\href{https://indico.in2p3.fr/event/10819/session/5/contribution/5/material/slides/0.pdf}{R. Kogler}$\,$\cite{this}
presented the indirect determination of $M_t$ from precision data:
\beq M_t = 177\pm 0.5_{\rm th}\pm 2.4_{\rm exp}\qquad   \hbox{with uncertainty dominated by $M_W,\sin^2\theta$.}\eeq
Similarly, a global fit of flavour data gives\footnote{Thanks to  Pierini, Silvestrini, Paradisi.}
\beq  M_t = 171.5 \pm 7 \GeV.\eeq
All these measurements agree with the more precise direct measurement of the top pole mass
performed at TeVatron and LHC:
\beq M_t=173.34\pm 0.76\GeV.\eeq
However, the top pole mass is only defined up to a non-perturbative QCD uncertainty of about $\pm\Lambda_{\rm QCD}\approx 0.3\GeV$.
Furthermore, this measurement of the
peak in the invariant mass of  final state particles produced by top decays
needs a significant amount of reverse engineering:
one sums over visible products, but MonteCarlo simulations are needed to rescale the result,
adding invisible neutrinos and removing initial state radiation.
Basically, this means employing a formula of the type:
$M_{\rm pig} =  K_{\rm MC} \sum M_{\rm sausages}$.
Data are now rich enough that the kinematical features predicted by MonteCarlo simulations can be validated by observations.

But still it's better if sausage-making is discussed among experts behind closed doors.
Two workshops discussed how to reduce the estimated theorethical uncertainty of about $\pm 0.3\GeV$.
The result was an increase in the estimated theoretical uncertainty up to $\pm1 \GeV$.
\href{https://indico.in2p3.fr/event/10819/session/6/contribution/83/material/slides/0.pdf}{S. Weinzierl}$\,$\cite{this} reported that QCD 
experts now know what to do 
to improve the situation with 
short-distance definitions of $M_t$
(somewhat analogously to what done in the past decade for defining jets).

\section{The quest for physics Beyond the Standard Model}\label{natsec}

The question asked by Dirac  --- why $F_{\rm gravity} \sim 10^{-40} F_{\rm electric}$? --- 
becomes in the Standard Model
 a deeper naturalness question 
about its only dimensionful parameter: why $M_h^2 \sim 10^{-35} M_{\rm Pl}^2$?
According to standard lore
``the Higgs mass receives power divergent quantum corrections
\beq\label{eq:nat}
 \delta M_h^2 \sim g^2_{\rm SM} \Lambda^2_{\rm UV}\circa{<} M_h . \eeq
New-physics must realise the cut-off
$\Lambda_{\rm UV}$ at the weak-scale 
such that $M_h$ is naturally smaller than the Planck scale''.

\href{https://indico.in2p3.fr/event/10819/session/9/contribution/111/material/slides/0.pdf}{G. Altarelli}$\,$\cite{this} recalled
early Moriond conferences: we heard that
supersymmetry as a solution to the naturalness problem started to be a major topic since 1982.
A review by H.P. Nilles already appeared in 1984, telling that
{\it ``experiments within the next 5{\color{red}-}10 years will enable us to decide whether
supersymmetry, as a solution to the naturalness problem of the weak interaction
is a myth or reality''}$\,$\cite{Nilles}.   

5-10 years later LEP measured the SM gauge couplings, finding values in agreement with SU(5)
unification in presence of  supersymmetry at the weak scale.

25 years later the theoretical community worried about the LHC Inverse Problem:
naturalness of the Higgs mass needs so much new sparticles 
at the electroweak scale that we will be inundated by new physics
and data might not be enough to unambiguously understand what it is.
LHC started, and the problem got solved:
the new sparticles predicted by supersymmetry
missed to show up at their appointment with history.

Even if they will come later, they might be too heavy for satisfying the naturalness requirement of eq.~(\ref{eq:nat}).

%
%

\begin{figure}
$$\includegraphics[width=0.65\textwidth]{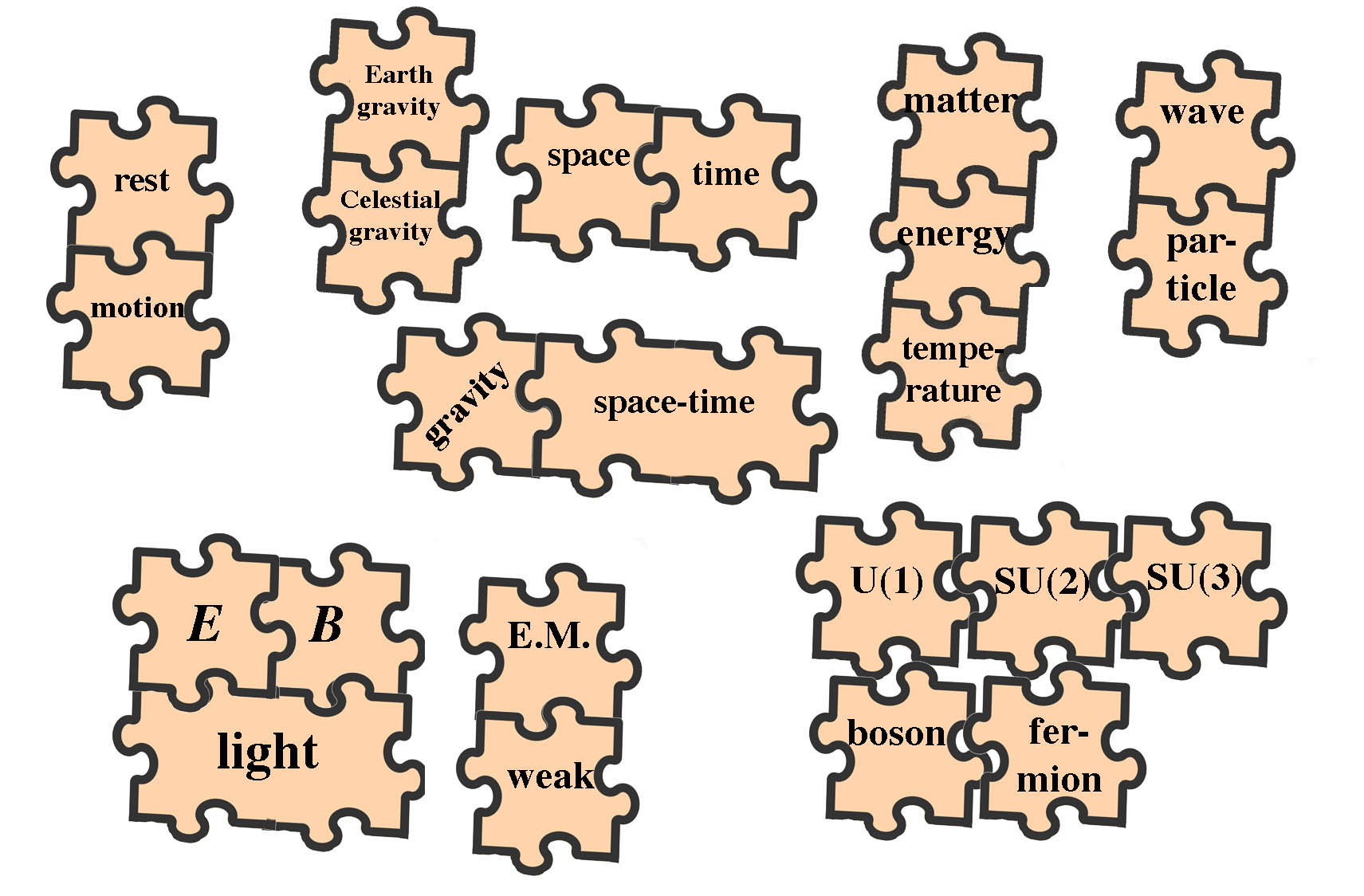}$$
\caption{Super-symmetry would continue the
path towards unification in physics.\label{unification}}
\end{figure}

\pagina{Is nature natural?}\noindent\label{natur}%
Super-symmetry is by far the best proposed natural theory, as it
 would continue the  path towards unification in physics
(see fig.~\ref{unification}):
supersymmetry allows for  SU(5), SO(10), E$_6$ unification of gauge couplings;
supersymmetry unifies bosons with fermions;
local supersymmetry is the gauge symmetry needed for a consistent theory of spin 3/2 particles,
supersymmetry avoids tachyon instabilities of strings...

These unification ideas can be tested experimentally:
gauge unification implies proton decay, and
weak-scale  supersymmetry gives observable signals at colliders.

That have not been observed.

The probability that a numerical accident made sparticles too heavy is
roughly given by the inverse of the fine-tuning.
Generic models where the Higgs mass is comparable to other soft-breaking SUSY parameters
(such as the Constrained MSSM) started to be fine-tuned after LEP.
Before LHC, this issue was seen as a motivation to search for
special natural models where the Higgs mass  is a loop factor lighter than other sparticles.
After LHC run I even these special models start to be fine-tuned.
Furthermore, within the MSSM, the measurement of the Higgs mass $M_h\approx 125\GeV$  points to even heavier sparticles.
Theorists specialised in naturalness model-building propose ways to reduce the fine-tuned $S$trangeness but at the price of
delivering ultra-special models with increased $B$aroqueness, without reducing
the $BS$ product.\footnote{The $BS$ plane was first introduced by A.\ Falkowski.}

From a theoretical point of view,
it is difficult to believe that supersymmetry could be the wrong path: 
one it tempted to hammer together the pieces of the puzzle, beliving
that  supersymmetry will be discovered and that
the quotation by Nilles can be fixed by just removing the dash between 5 and 10.
\href{https://indico.in2p3.fr/event/10819/session/10/contribution/76/material/slides/0.pdf}{A. Soni}$\,$\cite{this}
emphasised that we must be sure before abandoning naturalness:
exploring the 10-100 TeV range with flavour and collider experiments would allow to reach fine-tunings up to about $10^{4-5}$.
But colliders can no longer be bureaucratically justified by the warranty that 
some missing piece of the Standard Model will be discovered.
While the unnaturalness of the Higgs would be more interesting than the Higgs itself,
in the minds of funding agencies the exchange rate between Fine Tuning and Swiss Franc could not be good enough.




\pagina{Ant**pic selection in a multiverse}\noindent\label{ant}%
In the past, similar naturalness arguments correctly anticipated new physics
that keeps the electron mass naturally small (the positron),
that keeps the mass difference between charged and neutral pions naturally small (QCD compositeness),
that keeps flavor-violating effects naturally small (the charm).
In condensed matter systems,  the mass of higgs-like scalars that describe collective excitations
is naturally close to their ultimate cut-off (the Lorentz-breaking atomic lattice).

So, it is surprising that in particle physics naturalness might  not apply to the dimensionful parameters that today seem fundamental:
the Higgs mass, the Planck mass, the cosmological constant.

The observed cosmological constant,
$V_{\rm min} \approx 10^{-123} M_{\rm Pl}^4$, poses another naturalness hierarchy problem
that neither supersymmetry nor any other known theory can explain.
Then, anthropic selection in a multiverse of $>10^{120}$ vacua are invoked to justify the unnaturalness:
one of these vacua could accidentally have acosmological constant  small enough  that  galaxies can there form.
Then, observers can form and understand why the cosmological constant is so small that it starts to dominate just now.

\bigskip

The Higgs mass naturalness problem could also have an anthropic interpretation.
In the SM the hundreds of nuclei needed to develop a complex chemistry come out from a numerical accident:
the charged proton is the lightest baryon
(despite not being the most neutral baryon nor the baryon made of lighter quarks)
because these two competing effects are comparable:
\beq (y_d - y_u) v \approx \alpha_{\rm em} \Lambda_{\rm QCD}.  \label{proton}\eeq
From a fundamental point of view, the two contributions to the mass splitting 
could have differed by orders of magnitude.
If $v$ gets increased or reduced by more than about one order of magnitude, the proton
ceases to be the lightest baryon and complex chemistry disappears.
Then, many authors think that $M_h \ll M_{\rm Pl}$  is just another anthropic fine tuning.

\bigskip

Nobody talked about antrhopics at Moriond 2015.
This has an anthropic interpretation: Moriond is not in California.
Clearly, social factors are playing a role, as always when experiments cannot set the issue.
On one side, `having discovered the multiverse'  is physically indistinguishable from `having pursued a failed unification program',
but sounds much better.
On the other side, future physicists could consider us as crazy for not having immediately accepted anthropic arguments.




In my opinion, {\it anthropic selection in a multiverse does not explain an unnaturally small Higgs mass}.
The reason is that, even accepting that we live in some random vacuum in a multiverse, 
there is no need for a fine-tuning as unlikely as $v^2/M_{\rm Pl}^2$.
The most likely outcome should have been:
\begin{itemize}
\item one of the existing natural theories, like weak-scale supersymmetry;
\item or an anthropically acceptable alternative to the SM that does not involve an unnaturally light Higgs scalar; 
\item or, even within the Standard Model, more natural values of its parameters:
smaller $M_{\rm Pl}$ and bigger $v$ (compensated in eq.~(\ref{proton}) by reduced Yukawa couplings $y_{u,d}$).
\end{itemize}


\pagina{Subtle is the Lord...}\label{cart}
Today we are confused about naturalness, but
 nature is surely following some logic.
Suppose you get lost while traveling and encounter this signpost:
$$\includegraphics[width=0.4\textwidth]{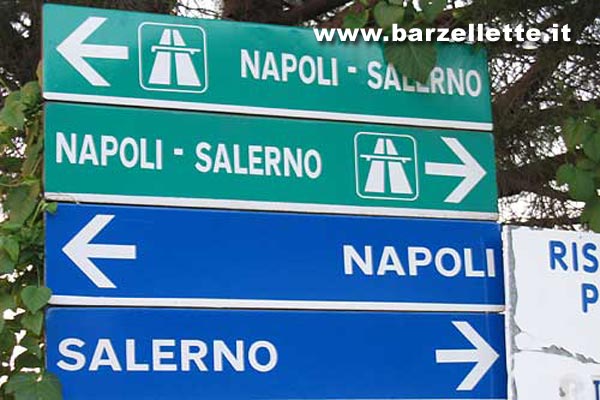}$$
What does it mean?
Unfortunately it's not clear. 
Fortunately there are theorists who start to think.
The smartest theorist immediately proposes a beautiful natural interpretation: 
there is some fundamental symmetry such that
\beq\hbox{Napoli = Salerno}.\eeq
However, after a bit of traveling, one starts to worry that this identification
might not be supported by geographical data.

\medskip

Then, another theorist
proposes an alternative anthropic interpretation:
$$\hbox{ it's just mafia selling signposts. }$$
Plausible, but how can it be confirmed?
After a debate about Popperian falsifiability and circular reasoning, 
a prediction emerges:
if there is mafia, then, probably, we are in Italy.
Correct, but probably you already knew it.

\medskip

To understand the real meaning one needs to think different --- 
in this case like the bureaucrats who place signposts.
Then, one can deduce where the photo was taken.
This is left as an exercise.

\medskip

\pagina{Data speaks and is telling SM, SM, SM}\noindent\label{jump}%
Summarizing:  the smallness of the Higgs mass seems unnatural (section~\ref{natur}), it does not seem antrhopic (section~\ref{ant}),
so  we should search for its hidden logic (section~\ref{cart}).

\smallskip

After the measurement of the Higgs mass
we finally have all the SM parameters.  We can now try to assume that the SM is valid up to unnaturally large energies
and see where this leads us to:
\begin{itemize}

\item []
{\bf Fact 1}: the SM can be extrapolated at least up to $M_{\rm Pl}$;

\item[]

{\bf Fact 2}: in the SM, $M_h \approx 130\GeV$ corresponds to $\lambda(M_{\rm Pl})\approx 0$;

\item[]
{\bf Fact 3}: in the SM, $\beta(\lambda) = d\lambda/d\ln E$ vanishes around $M_{\rm Pl}$.
\end{itemize}
For sure, these could be just accidents with no meaning; new physics can change these properties, by a bit or by a lot.
The goal of this discussion is not proofing anything, but searching for
possible messages in the data that we have now.
If this is the message from nature, it is incompatible with our ideas of naturalness.

\begin{figure}[t]
$$\includegraphics[width=0.46\textwidth]{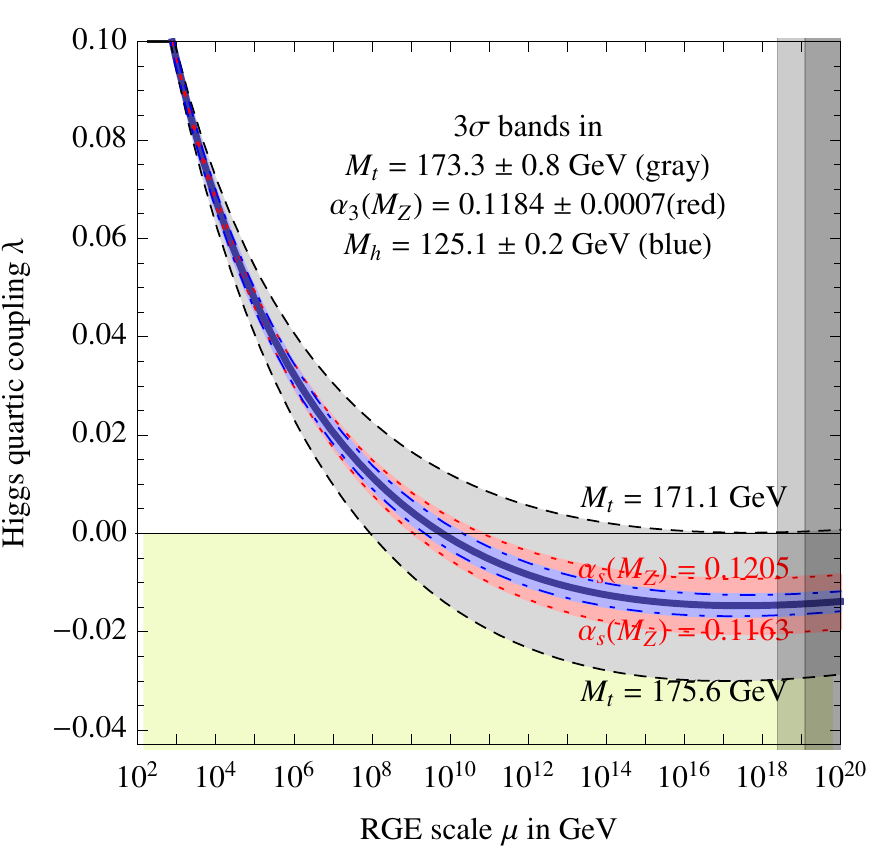}\qquad
\includegraphics[width=0.46\textwidth]{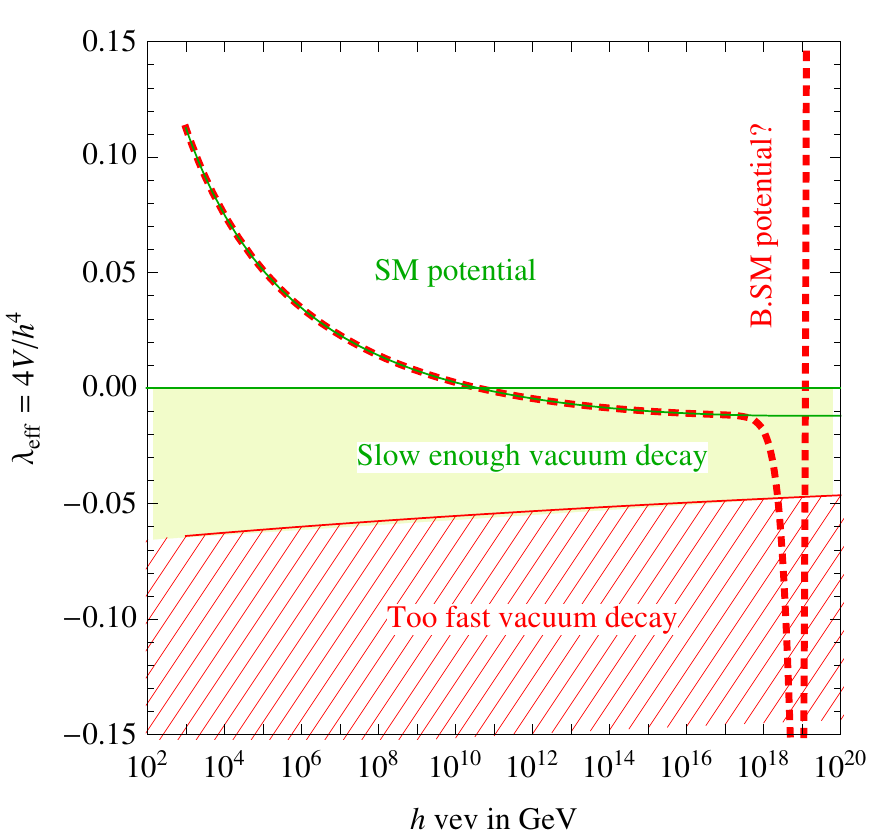}
$$
\caption{\label{lambda}Running of the Higgs quartic in the SM (left) and in the BranchinaSM (right).}
\end{figure}

\medskip


Looking more precisely at the SM, one notices that 
if $M_t \circa{>} 171\GeV$ and if the SM holds up to large energy,
the Higgs potential 
$V_{\rm SM} = \lambda_{\rm eff}(h) h^4/4$
is unstable, falling down at field values $h>h_{\rm max}$.
For present central values of the SM parameters,
the critical higgs field value is $h_{\rm max}\sim 10^{11}\GeV$, 
but uncertainties on $M_t$ can largely change it, see fig.~\ref{lambda}a.
This instability behind by a potential barrier  leads to vacuum decay,
with a rate proportional to $\exp(-8\pi^2/{3|\lambda_{\rm eff}|})$,
which is too fast if $\lambda_{\rm eff} < -0.05$.
The vacuum decay rate is safely small because
$\lambda_{\rm eff}$ gets negative but remains small.

\bigskip

\href{https://indico.in2p3.fr/event/10819/session/3/contribution/26/material/slides/0.pdf}{V. Branchina}$\,$\cite{this} 
showed that the vacuum decay rate can be much faster 
(possibly as fast as the rate at which he emphasises such statement)
in BSM models where $V_{\rm SM}$ is stabilised at $h \circa{<} M_{\rm Pl}$ in this peculiar way:
$$ V_{\rm BSM}  = V_{\rm SM} - \frac{h^6}{M_{\rm Pl}^6}+\frac{h^8}{M_{\rm Pl}^8}+\cdots$$
plotted in fig.~\ref{lambda}b.
In my opinion, searching for possible implications of data, it is better to focus on subtle possibilities, rather than on malicious ones.

\bigskip

The instability of the SM potential raises cosmological issues:
can cosmological evolution end up in the electroweak vacuum,
despite that it is much smaller in field space
than the Planck-scale vacuum?
The answer is: yes, if the top mass lies in the special range
$171\GeV\circa{<} M_t \circa{<}175\GeV$. 
Indeed, during inflation with Hubble constant $H= 2.5~10^{14}\GeV\sqrt{r}$
(where  $r\circa{<}0.1$ as found by the \href{https://indico.in2p3.fr/event/10819/session/1/contribution/3/material/slides/0.pdf}{Planck} collaboration$\,$\cite{this})
the Higgs $h$ can fluctuate acquiring a large random vev, $h \sim H$.
Provided that $\langle h^2\rangle$ is not too large,
during the thermal phase after inflation the Higgs acquires a thermal mass $\approx T$,
such that its vacuum expectation value returns to zero, 
finally landing in the SM minimum.
$M_t\circa{<}175\GeV$ is needed to guarantee  the happy end: otherwise
thermal tunneling can be too fast.

\href{https://indico.in2p3.fr/event/10819/session/3/contribution/117/material/slides/0.pdf}{A. Kusenko}$\,$\cite{this} showed how this out-of-equilibrium phase could provide baryogengesis
(altought extra ad-hoc model building is needed to avoid  iso-curvature perturbations,
which are disfavoured by cosmological observations).

%
%
%
%

\pagina{Crazy alternative ideas about naturalness}\noindent
Usual naturalness, as formulated in eq.~(\ref{eq:nat}),
attributes physical meaning to  regulators and to power divergences.
However these are unphysical computational tools.
Maybe we are over-interpreting quantum field theory, analogously to what happened
a century ago, when theorists over-interpreted Maxwell wave equations as implying 
the existence of a medium where light propagates.
Experimentalists failed to find such \ae{}ther, opening the road to crazy alternative ideas.

\smallskip

Maybe naturalness only demands that physical (potentially observable)
quantum corrections to the Higgs mass are naturally small.
Then the SM alone would be natural, for its observed values of the parameters.
New physics can also be natural provided that
\beq {\delta M_h \sim g_{\rm BSM} M_{\rm BSM} \circa{<} M_h}.  \label{modn}\eeq
This naturalness condition differs from the usual one, eq.~(\ref{eq:nat}), because
$g_{\rm SM}$ (the SM couplings of the Higgs boson, such as the top Yukawa coupling) is
replaced by couplings $g_{\rm BSM}$ to new heavy particles, and the cut-off
$\Lambda_{\rm UV}$ by $M_{\rm BSM}$, the mass of extra particles.
For example, SU(5)  gauge unification needs heavy vectors, not compatible with eq.~(\ref{modn}). 

\smallskip

My collaborator
\href{https://indico.in2p3.fr/event/10819/session/3/contribution/1/material/slides/0.pdf}{A. Salvio}$\,$\cite{this} 
argued that power divergences must vanish if nature is described, at fundamental level, by a theory with no dimensionful parameters.
Quantum corrections break classical scale invariance giving a renormalization group running for the dimensionless couplings,
allowing to dynamically generate mass scales.
He showed how the dimensionless  extension of Einstein gravity, agravity, 
is renormalizable and allows to dynamically generate the Planck mass, provided
that a scalar quartic runs in such a way that it and its $\beta$ functions simultaneously vanish around the Planck scale.
This is similar to how the SM Higgs quartic runs if $M_t \approx 171.1\GeV$, see fig.~\ref{lambda}a.
In agravity  the usual gravitational coupling  $g_{\rm gravity} \sim E^2/M_{\rm Pl}$ gets substituted
at large energy $E\circa{>}M$ by dimensionless couplings: naturalness, as redefined by eq.~(\ref{modn}), is satisfied provided that 
$M$ is low enough $M \circa{<} \sqrt{M_{\rm Pl} M_h}\sim 10^{12}\GeV$.
The new physics with mass $M$ involves a ghost-like particle, that might or might not receive a sensible
quantum interpretation.
Salvio also discussed inflationary predictions 
($n_s\approx 0.96$ and $0.003<r<0.13$) and attempts of finding natural
extensions of the SM that can hold up to infinite energy, avoiding any Landau pole.

\bigskip

Furthermore,
a recent paper claims to have, for the first time, a conventionally
natural model that does not need new physics at the weak scale, 
thanks to a cosmological evolution that selects a small Higgs mass$\,$\cite{GKR}.


\section{Conclusions}\noindent
Not today.  LHC run II is starting now, and
results presented at Moriond 2016 could bring us much closer to the conclusion.

\small

\section*{Acknowledgments}
I thank the organisers for creating a  unique environment for  interchange of ideas, and all participants  for a multitude of new insights.

%
%

\end{document}